# Thermal Adaptation in Viruses and Bacteria

Peiqiu Chen[1,2],   Eugene I. Shakhnovich[1]

[1]Department of Chemistry and Chemical Biology, Harvard University, 12 Oxford Street, Cambridge, MA 02138

[2]Department of Physics, Harvard University, 17 Oxford Street, Cambridge, MA 02138

# ABSTRACT

A previously established multiscale population genetics model states that fitness can be inferred from the physical properties of proteins under the physiological assumption that a loss of stability by any protein confers the lethal phenotype to an organism. Here we develop this model further by positing that replication rate (fitness) of a bacterial or viral strain directly depends on the copy number of *folded* proteins which determine its replication rate. Using this model, and both numerical and analytical approaches, we studied the adaptation process of bacteria and viruses at varied environmental temperatures. We found that a broad distribution of protein stabilities observed in the model and in experiment is the key determinant of thermal response for viruses and bacteria. Our results explain most of the earlier experimental observations: striking asymmetry of thermal response curves, the absence of evolutionary ''trade-off'' which was expected but not found in experiments, correlation between denaturation temperature for several protein families and the Optimal Growth Temperature (OGT) of their host organisms, and proximity of bacterial or viral OGTs to their evolutionary temperatures. Our theory quantitatively and with high accuracy described thermal response curves for 35 bacterial species using, for each species, only two adjustable parameters – the number of replication rate determining genes and energy barrier for metabolic reactions.

# INTRODUCTION

Temperature is one of the most important physical parameters in evolution. It defines a species' fundamental properties and plays an important role in many complex physiological mechanisms. Experimental observations have shown that temperature is essential in regulating the metabolic rate, the physiological changes of proteins, as well as the organism's mutation and replication processes. Many laboratory experiments have been carried out to study fitness response to stress caused by elevated environmental temperature (BENNETT and LENSKI 2007; BRONIKOWSKI et al. 2001; CULLUM et al. 2001; KNIES et al. 2006; TRAVISANO and LENSKI 1996) (BENNETT and LENSKI 1997; COOPER et al. 2001; HOLDER and BULL 2001; LEROI et al. 1994). There has been experimental evidence that the thermal niches for *E. coli* are asymmetrical between the high temperature end and at the low temperature end (CULLUM et al. 2001; TRAVISANO and LENSKI 1996). This means that while bacteria that have adapted to higher environmental temperatures can easily survive at lower environmental temperatures without sacrificing too much fitness, the fitness in general has a much sharper decline at the boundary of the higher temperature thermal niche (CULLUM et al. 2001) (TRAVISANO and LENSKI 1996). Various competition assays have also shown that culturing bacteria originating from the same strain in environments with higher temperatures, will bring them a greater competitive advantage on average than the original wild type strains, even when they are competing with the wild type strain at the original wild-type environmental temperature (BENNETT and LENSKI 1997; CULLUM et al. 2001; LEROI et al. 1994), in variance with common expectation of an evolutionary trade-off. In addition to laboratory studies, prokaryotes that have been isolated from high and low temperature environments suggest such asymmetries as well. Many of the prokaryotes that have been isolated from high temperature environments are obligate thermophiles, whereas those isolated from low temperature environments are not obligate psychrophiles and generally grow optimally at higher temperatures (KNOLL and BAULD 1989). Despite the abundance of experimental evidence, a quantitative and comprehensive explanation of the various thermal adaptation patterns in prokaryotes has been elusive. Although previous approaches (RATKOWSKY *et al.* 2005; XIA 1995) have brought considerable insight regarding the relationship between the environmental temperature, genome size and organism fitness, these models often use a relatively large number of adjustable parameters and sometimes fail to explain the fundamental connections between the asymmetric thermal adaptation behavior of an organism and the physical properties of their proteins. Therefore, based on our previous molecular evolutionary model (ZELDOVICH *et al.* 2007b) that all essential genes have to satisfy the minimal stability requirement for an organism's survival, we now present a model that can explain the adaptation of prokaryotes in a broad range of thermal environments. Our model explains the physical reason for the existence of the thermal niche asymmetry and lack of evolutionary tradeoff. It also defines a quantitative relationship between the number of proteins controlling the replication process in an organism, the free energy barrier on metabolic reactions, and the fitness response to elevated or decreased environmental temperatures.

**THE MODEL**

Our previously developed evolutionary model (ZELDOVICH *et al.* 2007b) provided certain insights into the distribution of the stabilities of all the essential genes in a genome. This model is based on recent experiments that showed that the knockout of any essential gene confers a lethal phenotype to an organism (FRASER *et al.* 2000; HERRING and BLATTNER 2004). Therefore, it assumes the fundamental minimalistic ''bare-bone'' genotype-phenotype relationship: *in order for an organism to be viable, all of its essential genes must encode (at least minimally) stable proteins*. This evolutionary model also assumes that protein stability is essentially a physiologically neutral trait, so long as the protein possesses sufficient stability to stay in the folded state (BLOOM *et al.* 2007). Based on this model, along with other sufficient experimental evidence about the protein stability distribution (KUMAR *et al.* 2006; SANCHEZ *et al.* 2006), we were able to provide a quantitative description of the stability distribution of all the essential genes within a certain genome.

In our model, we assume that the replication rate of an organism depends on the functionality of each of the proteins involved in the replication process of the organism. Replication does not necessarily involve all the proteins in the organism. Instead, it could be a smaller subset of the total number of proteins in the organism, and the number may vary among species and strains (NISHIKAWA *et al.* 2008). However, the organism is only able to replicate efficiently when all of these proteins are able to function properly (COUNAGO *et al.* 2006). Failure of a single rate-determining protein might result in the organism's dysfunction and hence reduce the organismal replication rate.

In our model an organism's genotype is essentially represented by the stabilities of all its proteins. Further, we focus on a subset of n proteins which control cell replication process and assume that replication rate of a bacteria or a virus is proportional to the copy number of each of the functional (i.e. properly folded) proteins which are encoded by ''rate-determining'' genes (see below). Here we note that the subset of rate determining genes may be smaller than the subset of essential proteins. The difference between the two is that essential genes (i.e the ones whose knockout causes lethal phenotype) may not affect growth rate directly. In contrast, the supply of functional proteins which are encoded by rate determining genes (below we will call protein products of rate determining genes rate determining proteins) may affect critically the ability of a cell to replicate. An example of such proteins could be enzymes involved in nucleotide and amino acid metabolism, DNA polymerases etc. Essential but not necessarily rate determining proteins may be enzymes which are involved in metabolism of certain nutrients, proteins responsible for motility under certain conditions etc.

We therefore denote $(\Delta G_1, \Delta G_2, ...\Delta G_i...\Delta G_n)$ as the organism's genotype, where $\Delta G_i$ represents the free energy of protein folding (i.e. the free energy difference between the folded and unfolded states) for the i$^{th}$ rate-determining protein in the organism, measured in kcal/mol. We can approximately treat the distribution of the stabilities of these n proteins as a continuous distribution.

It is generally accepted that an organism's metabolic reaction free energy barrier affects the rate of organismal replication (COUNAGO *et al.* 2006). Previously, many experiments (COUNAGO *et al.* 2006; RATKOWSKY *et al.* 2005) showed that organism replication rates have an Arrhenius-type dependence on temperature, with the slope (in proper variables) corresponding to the metabolic free energy barrier. The effect of the metabolic free energy barrier on birth rate can be expressed as:

$$b(\{\Delta G_i\}, T) \propto e^{-\frac{H}{k_B T}} \qquad (1)$$

Here $k_B$ is the Boltzmann constant, T is the environmental temperature, and H is the effective free energy barrier of the metabolic reactions. Although each species has its own effective metabolic reaction free energy barrier, experimental results showed that for most of the bacteria and viruses studied, H usually ranges from 10 kcal/mol to 20 kcal/mol (RATKOWSKY *et al.* 2005). When temperature decreases, the organisms will have a slower rate of metabolic reaction, and therefore, a slower rate of organism replication as well.

Rate determining proteins, like any other protein, are active only when folded (natively unfolded proteins were not found in viruses and bacteria). If any of the rate determining proteins loses its stability its copy number in the folded (i.e. functional) form will be reduced and the organism replication rate will drop as a result. Assuming that the expression levels of all rate-determining proteins are independent on temperature, we posit that for an organism with n rate determining genes, and the fraction $[f_i]$ of each rate determining protein in the folded state in the organism, the replication rate should be:

$$b(\overline{\Delta G_i}, T) \propto \prod_{i=1}^{n} [f_i] \qquad (2)$$

The simple form of the dependence of growth rate on protein stability suggested by Eq.2 is motivated by the view that in order for cells to function and duplicate, their major metabolic and biosynthetic pathways should be operational. Since many of them involve various proteins in sequential manner, the loss of copy number of any of them could result in a bottleneck-like effect on the total replication rate. The form of Eq.2 is similar in spirit (but not in detail) to the ''weak-link'' hypothesis on which recent successful model of early evolution was based (ZELDOVICH *et al.* 2007a). We can further define $[f_i]$ by quantitatively modeling the fraction of the folded proteins with folding free energy, $\Delta G_i$. Here we take into consideration that the folding of many protein domains is thermodynamically two-state, with only the folded and unfolded states being stable or metastable, and where $G_i^f$ and $G_i^u$ are the free energies of the folded and unfolded forms of protein i respectively. $\Delta G_i = G_i^f - G_i^u$ is the free energy difference between the two forms. To avoid confusion we note that each state – folded and unfolded – is viewed here as an ensemble of conformations corresponding to free energy minimum with respect to a relevant order parameter describing the degree of folding of a protein (SALI *et al.* 1994; SHAKHNOVICH 2006b). Thus, $\Delta G_i$ represents the stability of protein i. Therefore, for two-state proteins, the fraction of proteins that remain in their native state can be approximated as:

$$f_i = \frac{e^{-\frac{G_i^f}{kT}}}{e^{-\frac{G_i^f}{kT}} + e^{-\frac{G_i^u}{kT}}} = \frac{1}{1 + e^{\frac{\Delta G_i}{kT}}} \qquad (3)$$

It is clear from equation (3) that a lower value of the free energy of protein folding translates into a higher ratio of the folded proteins to unfolded proteins in the organism. The folding free energies of proteins depend on their sequences, i.e., on the genotype of the organism and environmental conditions (temperature, pH etc) (PRIVALOV 1979).

Therefore, the population growth rate b can be expressed as a product of the Arrhenius-type factor and the fraction of properly-functioning rate-determining proteins:

$$b(\overline{\Delta G_i}, T) = b_0 \frac{e^{-\frac{H}{k_B T}}}{\prod_{i=1}^{n}(1 + e^{\frac{\Delta G_i}{k_B T}})} \quad (4)$$

Because the proteins' folding free energy is a function of temperature, the folding free energy will change when temperature is perturbed. Therefore, for a given protein which has a folding free energy of $\Delta G_i$ at its original temperature T, its folding free energy $\Delta \tilde{G}_i$ will be different, if the environmental temperature is perturbed by $\delta T$.

In their native conformations, proteins possess low enthalpy and low entropy, whereas in the denatured (unfolded) state, their enthalpy and entropy are both high. Proteins unfold at temperature $T_F$, when the free energy of the native state, $G_F$, equals to that of the unfolded state, $G_U$:

$$\Delta G = G_F - G_U(E_U, S_U, T) = 0 \quad (5)$$

where $G_F = E_F$ (since $S_F \sim 0$), $G_U = E_U - k_B T_F S_U$.

Here, $G_U, E_U, S_U$ are the free energy, the enthalpy and the entropy of unfolded states, and $G_F, E_F, S_F$ are the free energy, the enthalpy, and the entropy of the folded states. We also assume here that the entropy of the folded state is small, and is therefore negligible compared to the non-native state entropy. The entropy of the unfolded state also does not change significantly with temperature (MAKHATADZE and PRIVALOV 1995; PRIVALOV 1979; SHAKHNOVICH and FINKELSTEIN 1989) in a relatively narrow range of temperatures which we study here. Assuming that the entropy of the unfolded state is approximately constant for an average sized 100-amino-acid domain (PRIVALOV 1979), the free energy when the protein unfolds can be expressed as:

$$G_F = E_U - k_B T_F S_U \qquad (6)$$

As changing temperature does not influence the energy of the protein very much, we can see from this equation that temperature would mostly influence a protein's folding free energy by placing a greater factor in front of the entropy term. Therefore, given an approximately constant $S_U$ and $E_U$, the $G_F$ will be smaller when temperature T is higher, and greater when temperature T is lower.

The total entropy difference (which also includes entropy of the solvent) between the folded and unfolded states is directly measured in thermodynamic experiments as the $\Delta S$ of denaturation ($\Delta S = S_U - S_F \cong S_U$) (MAKHATADZE and PRIVALOV 1995; PRIVALOV 1979; ZELDOVICH et al. 2007b). While $\Delta S$ varies with temperature in a broad temperature range (hence the phenomenon of cold denaturation (PRIVALOV et al. 1986)), experimental results show that in a narrower range of temperatures, around mesophiles' living temperature (20°C-40°C) the value of $k_B \overline{\Delta S}$ can be proaximated by a constant $0.25 kcal \cdot mol^{-1} \cdot K^{-1}$ (MAKHATADZE and PRIVALOV 1995; PRIVALOV 1979). Therefore, according to equation (5), for mesophiles, when temperature changes from T to $T + \delta T$, a protein will become less stable, having a higher folding free energy, which is approximately

$$\tilde{G}_i = G_i + k_B \overline{\Delta S} * \delta T = G_i + 0.25 * \delta T \qquad (7)$$

In most thermal adaptation experiments, the temperature change is immediate, and organisms do not have time to readjust stability of their proteins through acquisition of mutations during the temperature change. Therefore, upon temperature increase, the protein stability distribution $p(\Delta \tilde{G}_i)$ would preserve its original shape, however the free energy of each protein will be shifted by a constant value of $0.25 \delta T kcal/m$ shifting the whole distribution as illustrated in figure 1.

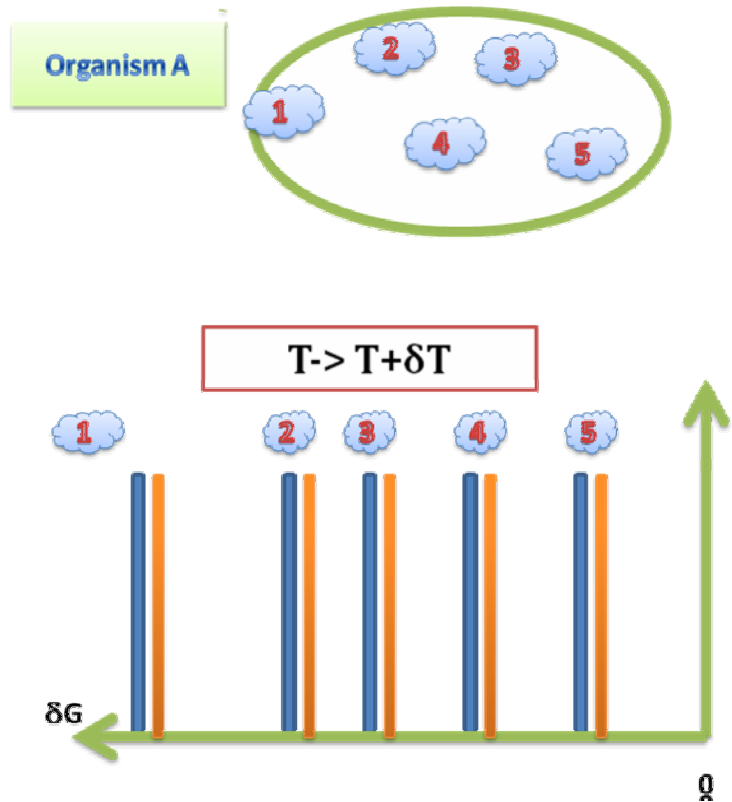

*Figure 1. An illustration of how the protein stability distribution changes upon an increase of environmental temperature. Blue bars: protein stability at initial temperature T; An illustrative organism A, has five rate determining genes in the cell, each with a different initial folding free energy. Upon temperature increase, each of the proteins becomes less stable, but the shape of the stability distribution is initially preserved.*

Thus, when temperature is increased, all of the proteins simultaneously become less stable, which will decrease the fraction of time that each of them stays in its native fold, thereby lowering the effective concentrations of functional form of these proteins.

From the analysis of the response of organismal fitness to thermal changes, we can see that temperature increase has a number of effects on the organism. First, because the metabolic reaction free energy is insensitive to temperature, the metabolic reaction rate will increase as temperature increases. Second, because each protein becomes less stable at higher temperatures, they each spend less time in their native states, and therefore function less efficiently as temperature increases. These two effects counter each other, and therefore, a more quantitative discussion is required in order to see which factors influence the fitness of the organism the most at any given temperature.

We have divided the following discussion into three sections. In the first section, we present results from numerical analysis for the thermal adaptation process, and we discuss the thermal adaptation behavior for bacteria and viruses with semi-conservative and conservative replication processes respectively. In the second section, we develop a semi-analytical model for the thermal adaptation process, and quantitatively discuss the relationship between the various parameters that are relevant for thermal adaptation and thermal response curves, as well as estimate the optimal growth temperature associated with each species. In the last section, we compare experimental results with the model predictions. We find that our model can provide a good explanation for the thermal adaptation of bacteria, using only two independent parameters for each bacterial species.

## RESULTS

### Simulation of Thermal Adaptation

We prepare the initial species with initial protein stability distribution drawn from the analytical distribution of the functional form described by (ZELDOVICH *et al.* 2007b) We then allow the species to evolve and adapt under the steady thermal environment for 20000 generations to equilibrate their protein stability distributions. (see Methods for details of simulations).

In our analysis of thermal adaptation responses, we also study the evolution and adaptation for conservative and semiconservative replication processes respectively. A semiconservative replication process means that during organismal replication, mutations can occur in both descendant copies, and this corresponds to many of the replication processes in DNA viruses and bacteria, which do not have methylation mechanism to distinguish between parent and newly synthesized strands. Conservative replication processes, on the other hand, describe the replication process for single strand RNA viruses, and bacteria which have methylation mechanisms to discriminate between parent and daughter strands i.e., when replication occurs, one copy (or strand) retains the same genome sequence as the previous generation, while the other copy (or strand) may mutate. Bacteria species usually have a much lower mutation rate than RNA virus species (DRAKE 1993; DRAKE et al. 1998; SNIEGOWSKI et al. 1997), which leads to a considerable difference in their thermal responses, as well as their thermal adaptation dynamics as will be shown below.

In our numerical study, for semi-conservative duplicated species, we chose a bacterium species with a metabolic free energy barrier of H = 20 kcal/mol, and n = 50 rate determining genes. For conservatively duplicated species, as RNA viruses usually have small genome sizes, we took the number of rate determining genes as n = 20, and chose a metabolic reaction free energy barrier of H = 20 kcal/mol. As will be explained in the comparison with experiment section, these are realistic parameters, since these values give thermal response predictions that agree well with experimental observations of the thermal adaptation behavior of mesophiles. Meanwhile, we chose the bacterium species to have a mutation rate of 0.003 mutations per genome per replication, as indicated by previous experiments (DRAKE et al. 1998), while the RNA viruses have a much higher mutation rate of 1.5 mutations per genome per replication, as experimentally found in some strains of polioviruses (DRAKE 1993).

After evolving in a steady thermal environment for 20,000 generations, the distribution of protein stabilities within an organism reaches equilibrium. We then study the response of populations to thermal shifts and compare the model results to experiments on thermal response and adaptation of bacteria and viruses (COOPER et al. 2001; CULLUM et al. 2001; HOLDER and BULL 2001; KNIES et al. 2006; LEROI et al. 1994; TRAVISANO and LENSKI 1996). We first varied the environmental temperature in order to observe the instantaneous organismal fitness response. This is consistent with experimental technique, where the fitness of bacteria and viruses is measured shortly after the environmental temperature change. Here we took the equilibrium species that evolved at 37 °C, and then varied temperature by altering the environmental temperature to T+$\delta$T, to determine the ratio of the new birth rate compared to the original birth rate. We study thermal response to change of temperature $\delta$T in the range from -15°C to 5 °C, and with step increment of 0.1°C.

On decreasing temperature, the fitness of both semi-conservatively and conservatively duplicated species underwent a slow decline, and on increasing temperature, their fitness underwent a sharp decrease. This occurred because after we equilibrated the species' protein stability distribution by evolving it for 20000 generation at fixed evolutionary temperature (37°C), the optimal growth temperature (OGT, the temperature at which the species reaches its maximum growth rate) appeared close to the evolutionary temperature of the species. In contrast, when temperature increases, the birth rates decline very rapidly, and will drop to around 20% to 50% of its original birth rate (the birth rate at 37 °C for this strain) within 5 °C of an increase. This is because when temperature increases, some proteins – least stable ones – significantly decrease their folded fraction in the organism, thereby decreasing the genome replication rate very rapidly.

In addition to the instantaneous thermal response, we also studied the long-time thermal adaptation processes of bacteria and viruses in our model in order to compare with long-time adaptation experiments. That is, we studied the thermal responses of bacteria and viruses after their adaptation to a new environment for a period of time. Experimentally, *E. coli.* strains that originally evolved at 37 °C were grown at other temperatures such as 42 °C or 20 °C for periods of time varying from a few months to 5 years (BENNETT and LENSKI 2007; LEROI *et al.* 1994). Experimental studies have revealed that these bacteria, after adapting to elevated environmental temperatures, possess improved fitness levels in a broad temperature range when compared to their ancestor strains. That is, for two *E. coli.* strains that are competing at 37 °C, the strain that has been previously cultured at a constant temperature of 42 °C for 5 years will have a higher fitness than the strain that has been growing under a constant 37 °C during the same 5-year period. Meanwhile, some experimental data also showed that although there is an elevated fitness for *E. coli.* strain that have been evolving at 42°C, the fitness improvement can sometimes be limited and not always significant. On the other hand, similar serial transfer experiments performed on viruses showed more significant thermal adaptation behavior compared to that of bacteria species(HOLDER and BULL 2001; KNIES *et al.* 2006).

In our simulation, we took the bacterium strain and the RNA virus strain that was initially evolved at 37 °C, increased temperature of its environment to 42 °C, and let the organisms evolve for a certain amount of time at this elevated temperature. Here in order to compare with experiments on *E. coli.* strains, we use 10000 generations, same with the experimental time scale for bacteria evolution. (For RNA viruses, because of their high mutation rates, we set the adaptation time as 1000 generations, and still were able to see a more significant adaptation than in bacteria, which agrees with experimental observation that viruses can exhibit more pronounced adaptation behavior (HOLDER and BULL 2001; KNIES *et al.* 2006).) We then measured the fitness as a function of the temperature change, where the fitness was measured as a fraction of fitness of the equilibrated wild-type at the original evolutionary temperature. After adapting the model bacteria at 42 °C for 10000 generations, we observed that the species had an elevated fitness level, even at its original temperature of 37 $^oC$. We also compared the protein stability distribution of the adapted strain to the wild type strain which was equilibrated at the initial evolutionary temperature of 37 °C. As shown in figure 2(B) and figure 2(C), the wild type strain had less stable proteins compared with the strain that evolved at elevated environmental temperature.

Meanwhile, from figure 2(A), we can see that even after 10000 generations at increased environmental temperature, the fitness improvement of the bacteria after culturing in 42°C environment is still relatively modest, especially when compared with that of the RNA virus (which only evolved for around 1000 generations). For RNA virus, whose mutation rates are much higher than bacterial ones, their proteins are less stable, and therefore more prone to thermal destabilization compared to that of DNA based organisms. However, as also discussed in the analytical calculation section of the paper, the effect of having more unstable proteins in RNA viruses is partially compensated by their short genome length. RNA viruses have much less rate determining genes than those of bacteria, therefore their OGT can still be lower than those of DNA based species (see next section for more explanation).

On the other hand, adaptation occurs faster in RNA viruses than for DNA-based organisms. When plotting the organismal fitness as a fraction of fitness value of equilibrated wild-type at their evolutionary temperature, we can see that RNA-based organisms have a more significant fitness increase when compared to DNA-based organisms, as shown in figure 2(A). Also, as can be seen from figure 2(B) and figure 2(C), after evolving at 42 °C for 1000 generations for the RNA virus and 10000 generations for the bacterium, the protein stability for the RNA virus is enhanced a lot more compared to that of the bacterium.

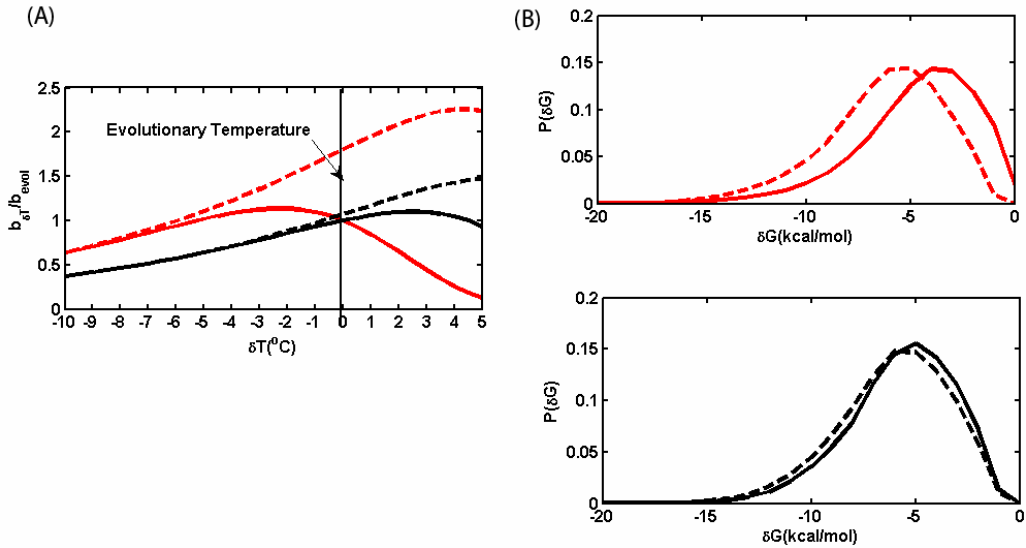

**Figure 2:** *Thermal response of fitness and protein stability distributions for a model bacterium species with 50 rate determining genes, mutation rate of 0.003 mutations per genome per replication (black curves)and model and a model RNA virus species with 20 rate determining genes, mutation rate of 1.5 mutations per genome per replication (red curves). Both species have been equilibrated for 20,000 generations at an evolutionary temperature 37 °C, and have a metabolic reaction free energy barrier H = 20 kcal/mol. For all four panels: Dashed line: strain evolved at 42 °C for 10,000 generations; Solid line: wild type strain evolved at 37 °C .* **Panel (A)**: *The fitness response at different temperatures.* **Panel (B),** *Protein stability distributionsfor the wild type (cultured at intial evolutionary temperature of 37 °C) for and cultured (at 42°C) strain of the RNA virus.* **Panel (C),** *Same as (B) for the bacterium species.*

In order to better understand the distribution of protein stabilities within each strain, we studied the denaturation temperature of all proteins for each strain . As discussed in the model section, as temperature increases, some of the proteins in the organism will become unstable and get denatured. The least stable proteins will get denatured first, while the more stable proteins will get denatured at higher temperature. Here we define the lethal denaturation temperature (LDT) for *an organism* as temperature above the evolutionary temperature where the least stable protein in this organism gets denatured, i.e its free energy $\Delta G = 0$. We plot the distribution of the organismal LDTs over all organisms in a population for each strain in Figure 3. It is clear from Figure 3, that highly mutating RNA virus populations form quasi-species since the distribution of their LDTs is broad and does not feature a pronounced peak. On the other hand, for the bacterium species, the distribution of LDTs is closer to a delta function, and, because they have more stable proteins, they also tend to have higher LDTs as well. From Figure 3 as well as from Table 1, we can see clearly that organisms from strains cultured at higher temperatures have higher LDTs, and the magnitude of increase for RNA virus is greater than that for the bacterium, because of the higher mutation rate in RNA virus, and thus more rapid and complete adaptation.

Meanwhile, we also study the mean denaturation temperature (MDT) for each strain, which is defined as the denaturation temperature (measured as deviation from the original evolutionary temperature of 37 C), averaged over all proteins in all organisms in a species. The result is listed as $T_{Den}^{mean}$ in Table 1 for each strain. From this analysis, we can see that although LDT for bacteria cultured at elevated temperature has improved significantly as can be seen in both Figure 3 and Table 1, MDT for cultured bacteria strain is however, not significantly different from the wild-type strain. This observation follows from the nature of the processes of mutation and selection process which occur during thermal adaptation. On the one hand, the selection pressure introduced by increasing the environmental temperature would eliminate organisms that contain very unstable proteins, thus the LDT of bacteria strain is significantly enhanced. On the other hand, the low mutation rate of bacteria strain, as well as the limited evolutionary time, gives the cultured strain limited opportunity to adapt to the new environment, which can also be seen from figure 2(C), that the stability distribution of all proteins in the population is not significantly different from each other, for the bacteria species. Thus at low mutation rates the adaptation process is essentially ''elimination of the least fit", while stabilities of proteins which remain stably folded even at elevated temperature is affected to a much lesser degree.

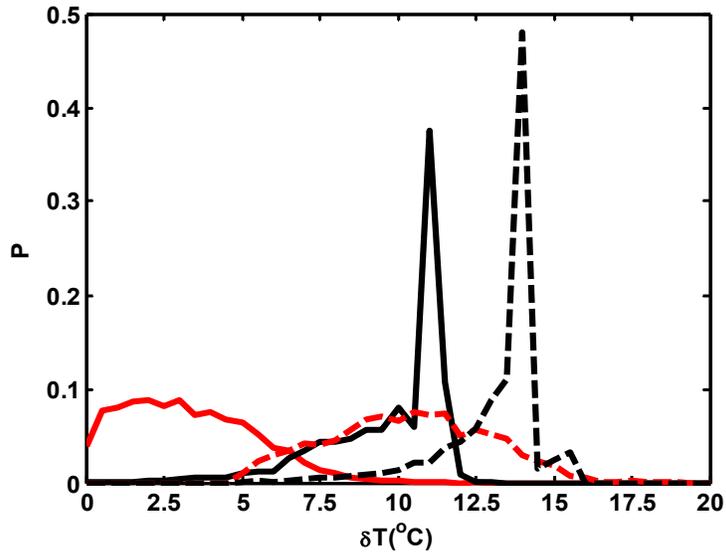

**Figure 3.** *Distribution of the Lethal Denaturation Temperature for organisms belonging to the four strains. Red Line: RNA virus; Black line: Bacterium; Solid line: wild type strains; Dashed lines: strains evolved at 42 °C;*

*Table 1*. *Protein denaturation temperatures for all four strains. Minimum Denaturation $T_{Den}^{\min}$ (°C) is the mean (over all organisms in the population) LDT in each strain; Mean Denaturation $T_{Den}^{mean}$ (°C) is the average temperature above the evolutionary temperature for each protein in the population to denature. All temperatures are measured as deviation from the original evolutionary temperature of 37 C*

|  | Minimum Denaturation $T_{Den}^{\min}$ (°C) | Mean Denaturation $T_{Den}^{mean}$ (°C) |
|---|---|---|
| Bacterium Wild Type | 9.7 | 27.6 |
| RNA virus Wild Type | 3.3 | 17.6 |
| Bacterium Cultured | 14.1 | 28.0 |
| RNA virus Cultured | 10.3 | 24.7 |

**Semi-analytical model for thermal adaptation**

To get better understanding of how various mechanisms described in our model influence the fitness of strains at different temperatures, we now apply mean field approximation, and calculate the dependence of fitness on the number of rate determining genes n, as well as the metabolic reaction free energy barrier H.

According to (ZELDOVICH *et al.* 2007b), the probability distribution of proteins' folding free energies within an organism, $\Delta G_i$, denoted as $p(\Delta G)$, can be approximately expressed as:

$$p(\Delta G) = C_0 e^{-\frac{h\Delta G}{h^2+D}} \sin\left(\pi \frac{\Delta G}{\Delta G_{max} - \Delta G_{min}}\right) = C_0(-e^{\frac{\Delta G}{\Psi}} Sin[\pi \frac{\Delta G}{L}]),$$

$$\text{for } -L < \Delta G_i < 0 \qquad (8)$$

Here, $C_0$ is the normalization constant for the probability distribution, D, h and L are parameters obtained from the distribution of energetic effects of the proteins' point mutations. D is the variance of $\Delta\Delta G$ change for a point mutation, h is the mean of the $\Delta\Delta G$ value for a point mutation. $L = \Delta G_{max} - \Delta G_{min}$ is the total range of viable free energies of protein folding and $\Psi = \frac{h^2 + D}{h}$. At room temperatures, $h \approx 1(kcal/mol)$; $D \approx 3(kcal/mol)^2$; and the value of $L = \Delta G_{max} - \Delta G_{min}$ is approximately 20 kcal/mol. The lower limit of the free energy distribution comes from the sequence depletion (MEYERGUZ *et al.* 2004; SHAKHNOVICH 2006a; SHAKHNOVICH 1998). According to the general theory of protein design (ENGLAND and SHAKHNOVICH 2003; SHAKHNOVICH 2006a; SHAKHNOVICH 1998) the ''inverse folding code'' is degenerate – i.e. there is a multitude of sequences which can have a given protein structure as their native state with particular stability $\Delta G$. The number of such sequences falls off rapidly as $\Delta G$ decreases (i.e. sequences form more stable proteins). Finally at a particular threshold value $L = \Delta G_{depletion}$ the sequence repertoire get exhausted forming the natural lower boundary for possible folding free energies of a protein (ZELDOVICH *et al.* 2007b). Experimental evidence shows this threshold value is around -20 kcal/mol at room temperature (KUMAR *et al.* 2006; SANCHEZ *et al.* 2006). While the distribution of stabilities given by analytical expression Eq.8 is slightly different from more accurate distributions obtained from simulations (Fig.2) it can be used as a reasonable first approximation as qualitatively (and semi-quantitatively) it captures the most features of experimentally observed distribution of protein stabilities (ZELDOVICH *et al.* 2007b)

From Equation (4), the organismal replication rate can also be rewritten as:

$$\ln b(\overline{\Delta G_i}, T) = \ln b_0 - \frac{H}{k_B T} - \sum_{i=1}^{n} \ln(1 + e^{\frac{\Delta G_i}{k_B T}}) \qquad (9)$$

Because the organism replication rate b is a function of the organism's genotype $(\Delta G_1, \Delta G_2, ... \Delta G_i ... \Delta G_n)$, from Equation (8), we also know the approximate probability distribution for the stabilities of the organism's rate determining proteinss. Given this information, it is convenient to take the mean-field approximation for the organismal birth rate and consider the ensemble average of all organisms in a species. In this way, we can calculate the average value of the organismal replication rate for a given species, and the summation over all n rate determining genes s can be approximated by integrating over the entire $P(\Delta G)$ distribution range. Therefore, the logarithmic population growth rate can be expressed as:

$$<\ln b(T)> \ = \ \ln b_0 - \frac{H}{k_B T} - n \int_{-L}^{0} \ln(1 + e^{\frac{\Delta G}{k_B T}}) p(\Delta G) d\Delta G \qquad (10)$$

The probability distribution $p(\Delta \tilde{G}_i)$, after an instantaneous temperature increase, can be expressed as:

$$p(\Delta \tilde{G}_i) = C_0 (-e^{\frac{\Delta \tilde{G}_i - k_B \overline{\Delta S} \delta T}{\Psi}} Sin[\pi \frac{\Delta \tilde{G}_i - k_B \overline{\Delta S} \delta T}{L}]) \quad \text{for} \quad -L + k_B \overline{\Delta S} \delta T < \Delta \tilde{G}_i < k_B \overline{\Delta S} \delta T \qquad (11)$$

Thus with the full analytical expression of the growth rate integral above, using equation (8)~(11), we can expand $<\ln b(T + \delta T)>$ to the second order of δT in the form:

$$<\ln(\frac{b(T+\delta T)}{b(T)})> \ = \ \delta T(nC_1+C_2)+\delta T^2(nC_3+C_4) \quad (12)$$

Here, $C_1 \sim C_4$ are various constants determined by H, the free energy barrier on metabolic reactions, T, the original growth temperature, and L, the total viable stability range (see Supplementary Information for their derivation). The number of genes determining the replication process in a genome is given by n. Since we know the functional form of the organismal birth rate and the analytical expression of the organism's protein stability distribution, we can calculate each of the coefficients $C_1 \sim C_4$ accordingly.

It is worthwhile to note that $C_2$ and $C_4$ come from the expansion of the metabolic free energy barrier term, and have the values $C_2 = \frac{H}{k_B T^2}$ and $C_4 = -\frac{H}{k_B T^3}$. Since $C_1$ and $C_3$ come from the expansion term with the protein stability distribution integral, using the property that $L \gg k_B T, L \gg \Psi$, we asymptotically get $C_1 \sim -\frac{k_B \overline{\Delta S} * k_B T}{(\Psi^2 + (k_B T)^2)}$ and $C_3 \sim -\frac{(k_B \overline{\Delta S})^2}{2(\Psi^2 + (k_B T)^2)}$ (See Supplementary Information for a more detailed analysis).

From these results, since $\Psi$ and $k_B T$ are both species-independent values, we can see that in biological systems, the species-dependent thermal adaptation behavior is largely controlled in our model by the metabolic reaction free energy barrier for a given species and the number of rate determining genes for that species.

**The Optimal Number of Rate-Determining Genes: Maximal Robustness to Temperature Fluctuations.** According to the expansion of the birth rate integral, we have the birth ratio relation for mesophiles living at room temperature (around 25°C):

$$<\ln(\frac{b(T+\delta T)}{b(T)})> \ = \ \delta T(nC_1+C_2)+\delta T^2(nC_3+C_4) \quad (13)$$

When the number of proteins involved in the replication process satisfies the relation $nC_1+C_2=0$, we denote this number as $n_c$, where at $n=n_c$, introducing a temperature perturbation $\delta T$ can only change the growth rate to the second order of $\delta T$. The implication of this for biological systems, is that when the environmental temperature fluctuates, due to either seasonal changes or species migration, the organismal fitness would not be greatly affected, and there would not be a drastic population expansion or decrease when the rate determining gene number is at $n_C$. In other words species having $n_C$ rate determining genes are most robust to temperature fluctuations. We can also show from the parameters that:

$$n_c \sim \frac{1}{T}\frac{H[\Psi^2+(k_BT)^2]}{k_B\overline{\Delta S}(k_BT)^2} \quad (14)$$

At $n=n_c$, introducing a temperature perturbation $\delta T$ can only change the growth rate to the second order in $\delta T$:

$$\ln(\frac{b(T+\delta T)}{b(T)}) \propto (\frac{\delta T}{T})^2 \quad (15)$$

From the equation above, we can see that for mesophiles living at close to room temperature, the optimal rate determining gene number, $n_c$, is mostly determined by H, the metabolic reaction free energy barrier of the organism for genome replication. When H is in the range 10 - 20 kcal/mol, and the growth temperature is around 25°C, $n_c$ can be 10-20 for mesophiles. We also acknowledge that the analytical description for the protein stability distribution is more concentrated around the lower stability end than the experimental database result (while the numerical simulation results of P(ΔG) shown in Figure 2 are in better agreement with experimental distribution at lower stability end). Therefore, substituting $P(\Delta G)$ in Eq. (10) with experimental protein stability distributions will give higher $n_C$ values, up to 40-60 as can be seen in Supplementary Table 1. For organisms with rate determining gene numbers $n$ that are greater than $n_c$, increasing temperature by a small amount $\delta T$ will decrease the birth rate. On the other hand, if $n$ is less than $n_c$, increasing temperature by a small enough $\delta T$ might modestly increase the fitness.

**The Optimal Growth Temperature:**

From the analytical expansion form of the thermal response of the organism, we can also see that there exists some critical temperature, $\delta T_C$, that satisfies the relation:

$$\partial_{\delta T} \frac{b(T+\delta T)}{b(T)} = 0 \qquad \frac{\delta T_C}{T} = \frac{-(nC_1/T + C_2/T)}{2(nC_3 + C_4)} \qquad (16)$$

At this temperature, the growth rate of this species will reach its maximum value. We therefore define $T + \delta T_C$ to be the optimal growth temperature for the species. From the analytical expressions of $C_1...C_4$, we can write that,

$$\frac{\delta T_C}{T} = \frac{H[\Psi^2 + (k_B T)^2] - n\overline{\Delta S}(k_B T)^3}{2H[\Psi^2 + (k_B T)^2] + n\overline{\Delta S}^2 (k_B T)^3} \qquad (17)$$

As we know, at T~300K, $\Psi \sim 3 kcal/mol$, $k_B T \sim 0.6 kcal/mol$, $\overline{\Delta S} \sim 125$. Therefore, from equation (17), depending on the specific value of the metabolic reaction free energy barrier and the rate determining gene number for each species, $\frac{\delta T_C}{T}$ is small. In fact, in the range of realistic values, where H ranges from 10kcal/mol to 20kcal/mol, n ranges from 10 to 50, and for T ~ 300K, $\left|\frac{\delta T_C}{T}\right| \in (2.4*10^{-3}, 7.4*10^{-3})$, which makes the OGT within three Celsius of the evolutionary temperature, for mesophile evolutionary temperatures.

This estimation shows that the optimal growth temperature (OGT) for a well-evolved mesophile species, should be quite close to the evolutionary temperature, which we assume to be its natural growth temperature. This is confirmed by many experimental observations (CULLUM *et al.* 2001; LEROI *et al.* 1994; TRAVISANO and LENSKI 1996).

We plotted the fitness change versus temperature for bacteria with different rate determining gene numbers, as well as for different values of the metabolic reaction free energy barrier, using our analysis of the semi-analytical calculations. We note that at constant metabolic free energy barrier and environmental temperature, increasing the number of rate determining genes, leads to lower optimal growth temperature. Analogously, the higher the value of n is, the more drastically the growth rate drops when the environmental temperature is increased, therefore, it becomes more difficult for an organism to adapt to elevated environmental temperatures.

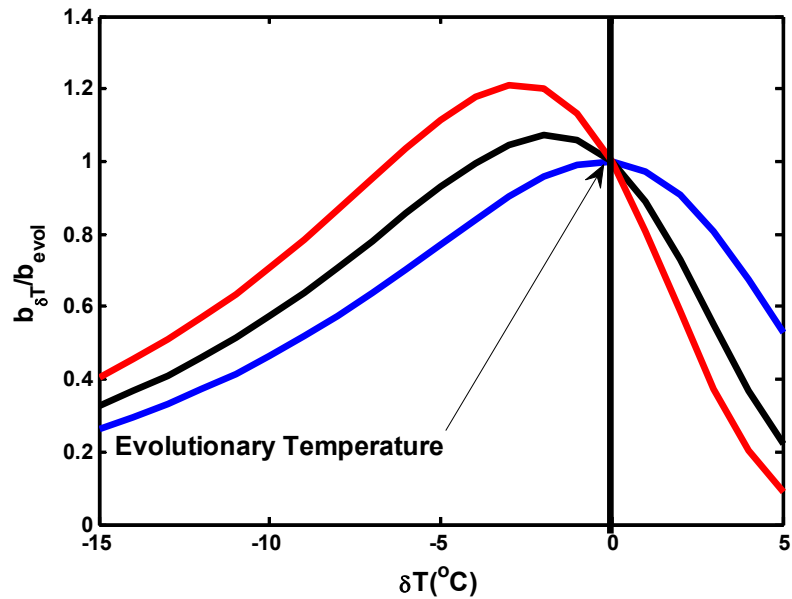

**Figure 4:** *The growth rate as a function of the value of $\delta T$ for different rate determining gene numbers, n. Red line: n = 30, black line: n = 20, blue line: n = 10. The growth rates*

*are measured as the ratio to the species replication rate at their original evolutionary temperature ($\delta T = 0$)*

For species that have different metabolic reaction free energy barriers, H, but with the same rate determining gene number n = 20, their birth rate ratio has been graphed in Figure 5 below (T is fixed at 27°C). We can see from the figure that the greater the metabolic reaction free energy barrier is, not surprisingly, the faster the birth rate decreases with decreasing temperature. On the other hand, according to equation (7), the fitness declines more slowly with increasing temperature when H is greater. This can also be seen in Figure 5, although this effect is relatively weak.

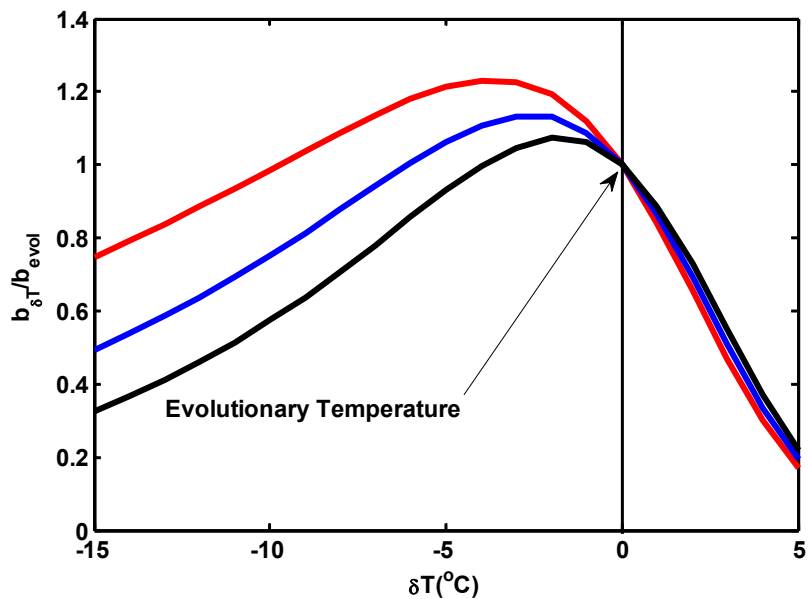

**Figure 5.** *Dependence of the growth rate on temperature for organisms whose genome sizes are at the optimal gene number. Red line; H = 10, blue line; H = 15, Black line; H = 20. T = 37°C and n = 20 for all three species.*

**Comparison with Experimental Results**

Several experiments have been carried out to study the thermal response and adaptation behavior of bacteria and viruses (BENNETT and LENSKI 2007; BRONIKOWSKI *et al.* 2001; RATKOWSKY *et al.* 2005). Ratkowski et al., have systematically studied 35 sets of data for thermal adaptation of different bacteria strains. Here, we analyzed these 35 mesophilic strains using our thermal adaptation model. Assuming that the metabolic reaction free energy barrier H, and the number of rate determining genes n are the only independent parameters for each strain, we fit the 35 datasets of bacterial thermal response with theoretical formulae. Since we have limited information about the evolutionary temperature for each bacterial strain, we used the OGT as a proxy for evolutionary temperature, motivated by the observation and our result that the two are not too different for mesophiles. We evaluate growth rate as function of temperature for each bacterial strain from Eq.10 using for P($\Delta$G) the distribution for experimentally measured stabilities of proteins derived in (ZELDOVICH *et al.* 2007b) from ProTherm database (KUMAR *et al.* 2006), Here we also consider instantaneous temperature change, upon temperature increase, whereby protein stability distribution P($\Delta$G) has the original shape but is shifted upon temperature change as explained above. Eq.10 contains two parameters – number of rate determining genes n and metabolic free energy barrier H which we adjust for each strain and checking a'posteriori that the values of these parameters are biologically reasonable..

We used the nonlinear regression method to find the H and n values associated with each strain, and for 35 independent bacterial strains, the rate determining gene number ranged from 10 to 50, which are roughly 10% to 30% of the number of the essential genes in the bacteria, thus representing a reasonable order of magnitude estimate for the number of rate determining genes in a specie. The metabolic reaction free energy barrier ranges from 10 to 20 kcal/mol, and this also agrees with previous estimates. Here, we were able to obtain a relatively good fit for almost all of the datasets, and several examples comparing the experimental data with our theoretical predictions are shown in Figure 6 below (see Supporting Information for fits for remaining 32 strains.).

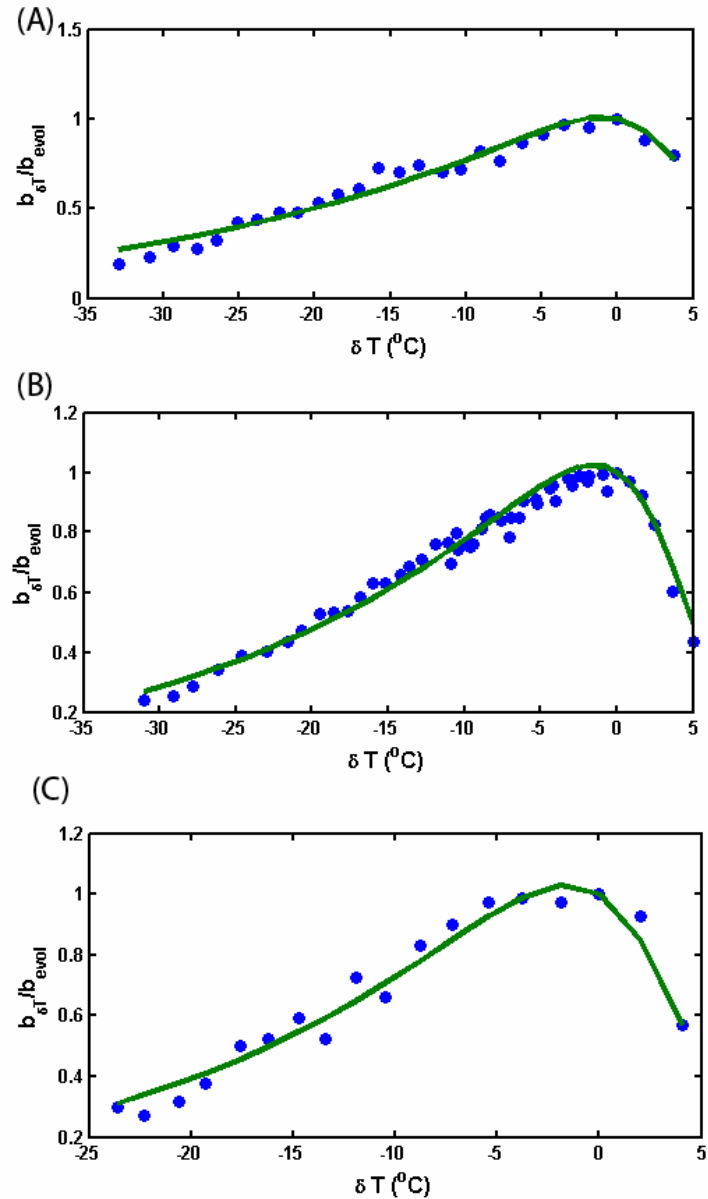

**Figure 6.** *Comparison between experimental thermal response curves and predictions from the semi-analytical model. Here we show results for 3 of 35 species studied by Ratkowsky and coauthors (RATKOWSKY et al. 2005). (a) L. monocytogen,H = 7.64 kcal/mol,*

*n = 23. (b) ps florescence.(H = 8.05kcal/mol, n =34 ), and (c)   E. coli.(H = 12.1 kcal/mol, n = 41).*

From Figure 6, we can see that by varying just two parameters, the number of genes controlling the thermodynamic process of replication and the genome replication free energy barrier, we can get a correlation of 89% to 99% between the experimental data and the analytical model prediction. Therefore, we can state that our model accurately describes the thermodynamic response of the adaptation process of bacteria. A further advantage of our model is that it contains far fewer parameters compared to earlier models, such as those in Ratkowsky et al. (RATKOWSKY *et al.* 2005), while being equally effective in describing experimental phenomena.

## DISCUSSION

Thermal adaptation in viruses and bacteria was studied extensively in the past, and a number qualitative features of thermal response and adaptation were found to be universal, common to all studied species and strains. In particular, the following observations were made: 1) that an OGT of an organism is very close to its evolutionary temperature (CULLUM *et al.* 2001; KNIES *et al.* 2006; LEROI *et al.* 1994; TRAVISANO and LENSKI 1996), 2) a pronounced asymmetry of thermal response curves in viruses and bacteria whereby their growth rate declines slowly with decreasing temperature while it declines more rapidly upon temperature increase (COOPER *et al.* 2001; KNIES *et al.* 2006) 3) lack of evolutionary trade-off whereby bacteria and viruses cultivated at higher temperature appear to be more fit than bacteria cultivated at original evolutionary temperature in a broad temperature range, including original evolutionary temperature (CULLUM *et al.* 2001; HOLDER and BULL 2001; KNIES *et al.* 2006) 4) Correlation between OGT of an organism and denaturation temperature for some   protein families.

Our model, while quite minimalistic, explains all these findings providing a unified picture of physical mechanisms of thermal adaptation. The key premise of the theory is that, in order to function, proteins must be stable, and that one of the key determinants of the rate of growth (i.e. fitness) of an organism is the amount of functional (i.e. folded) rate determining proteins available in the cell. Protein stability factor affects replication rate through modulation of the fraction of correctly folded proteins, as suggested by Eq.2 While Eq.(2) is empirical it is biologically justified in the sense that it assigns equal importance to the stability of each rate-determining proteins, while an alternative form where replication rate is proportional to the total copy number of folded rate-determining proteins would overweight the importance of highly expressed proteins and ignore the role of rate determining proteins which are expressed in lower copy numbers (e.g. some transcription factors and DNA polymerases). Nevertheless the dependence of replication rate on the copy number of folded rate determining proteins given by Eq.(2) is just a first approximation and other forms (which, e.g. put emphasis on toxic effect of misfolded proteins in the cells (DRUMMOND and WILKE 2008)) are possible and will be explored in future work.

The key novel aspect of our model is that it explicitly takes into account (and derives) a broad distribution of protein stabilities in the genome of a bacterial or viral species, in contrast to earlier studies which assumed that stability of a single protein determines the growth rate of bacteria (or that all proteins in an organism have the same stability) (RATKOWSKY *et al.* 2005). While the study of Ratkowsky and coauthors (RATKOWSKY *et al.* 2005) was successful in fitting thermal response curves for many bacterial strains such fitting had been achieved at the expenses of large number – 5 – of fitting parameters describing thermodynamics of the single rate determining protein. The fact that proteins stabilities are broadly distributed, determines many key features of prokaryotic thermal response.

The analytical approximation (Eq.16) and simulations show that deviation of the OGT from evolutionary temperature is small, in agreement with experimental observations. In (TRAVISANO and LENSKI 1996) the authors systematically studied the thermal response of *E.coli* after evolving in steady $37^oC$ environment for 20,000 generations. This evolved strain shows a direct and clear trend of fitness decrease as temperature deviates from the original $37^0C$. That is, the OGT appeared less than 1 °C different from its evolutionary temperature for this well-evolved *E. coli* strain. On the other hand, the ancestor strains, whose protein stability distributions may not be fully equilibrated within the population, show some small fluctuation of growth rate when the environment temperature increases by less than 2°C from its evolutionary temperature, eventually the fitness declines sharply upon further increase in temperature. Our theory provides the physical rationale for this observation. Indeed the broad equilibrium distribution of stabilities of rate determining proteins (see Fig.2B and 2C) implies that there exist ''weakest links'', i.e. the least stable proteins for which even a slight increase of temperature results in a significant decrease of equilibrium population of their folded form. It is the drop of the copy number of these *folded* rate determining proteins which brings about an immediate loss of fitness upon an increase of temperature above evolutionary temperature. On the other hand, the optimal growth temperature has to remain high enough so that the metabolic reaction free energy barrier does not significantly slow down the growth of the population. This effect is more significant when metabolic reaction free energy barrier H is high. Thus, as shown quantitatively in our model (Eq.(16)), the OGT is close to their evolutionary temperature in species that have large values for their metabolic reaction free energy barrier.

Thermal adaptation experiments have shown that *E. coli* has a free energy barrier for the metabolic reaction rate of approximately H = -14.3 kcal/mol (RATKOWSKY *et al.* 2005). Our theory also gives best fit H value of around 14 kcal/mol for many strains (as can be seen in Supplementary Table 1), while the best fit number of rate determining genes in *E. coli* is 41. This number is around 15% of its essential gene number, and thus may be a reasonable estimate. Then according to the analysis of equation (16), OGT of well-evolved *E. coli* should be within 2°C of its evolutionary temperature, which agrees with experimental observations.

The fact that the OGT is especially close to evolutionary temperature for the equilibrated species points out to an interesting evolutionary observation. Indeed, one can argue that thermodynamic ''vulnerability'' of least stable proteins to an increase of temperature may create an evolutionary pressure to make them more stable. However we do not observe that in this model and as an implication, experiment shows that the OGT is indeed very close to evolutionary temperature. The reason why there is no apparent pressure to stabilize least stable proteins is that as bacteria evolve at highly controlled environment (constant T) it optimizes its fitness at this particular environment without concern about adaptation in a different environment which the strain have not encountered during long evolutionary equilibration. On the other hand, evolutionary optimization of distribution of protein stabilities beyond an optimal one at a given environment does not improve fitness in this fixed environment. Thus our model shows how long-time equilibration evolves ''specialist'' bacteria which may be poorly adapted to challenges beyond the conditions it was exposed to. The reason is not that specialization is an advantageous trait but that it is the easiest evolutionary solution at a given environment and long-time equilibration helps to find this solution.

In the numerical study of thermal adaptation, we provided a quantitative insight of how viruses and bacteria respond to temperature changes. Our simulation results agree with those from previous experiments on the asymmetry of thermal response for different bacteria and viruses in that fitness exhibits a slow decline upon temperature decrease, and a sharp decline upon temperature increase (BENNETT and LENSKI 1997; BENNETT and LENSKI 2007; BRONIKOWSKI *et al.* 2001; COOPER *et al.* 2001; CULLUM *et al.* 2001; ELENA *et al.* 2007; KNIES *et al.* 2006; LEROI *et al.* 1994; RATKOWSKY *et al.* 2005; TRAVISANO and LENSKI 1996). The reason for such an asymmetry is that different factors affect fitness at lower and higher temperatures, with partial unfolding of the least stable proteins being the key factor decreasing fitness at high temperature. The OGT for the subset of least stable proteins is not far from their denaturation temperature, and, the fraction of an ensemble of proteins which are in the folded state is most sensitive to temperature near the mid-folding transition for two-state proteins. For that reason fitness drops steeply when temperature increases above the OGT.

Our results show no evolutionary ''trade-off'' in thermal adaptation, in agreement with many experimental studies. Several authors (BENNETT and LENSKI 1997; BENNETT and LENSKI 2007; CULLUM *et al.* 2001; HOLDER and BULL 2001; KNIES *et al.* 2006; LEROI *et al.* 1994) showed that bacteria and viruses that have adapted to elevated environmental temperatures, acquire fitness which is superior to strains that have been growing at the original temperature for an extended period of time, even when they are competing in the same *original* thermal environment. Evolutionary "trade-off" in thermal adaptation, whereby bacteria cultivated at higher temperature should have lower fitness at normal temperature than bacteria cultivated at normal temperature (COOPER *et al.* 2001; KNIES *et al.* 2006), has been expected to determine thermal response in bacteria, in variance with actual observations. The basis of the ''trade-off''' expectations lies in the widely held opinion that in order to function proteins must be not too stable to allow for function-related flexibility (DEPRISTO *et al.* 2005). Several arguments are usually presented in support of his view. First, it is argued that that stability of real proteins is not too high, hence there must be some tradeoff between stability and functionality (DEPRISTO *et al.* 2005). Second, experimental observations that some mutations in active sites of several enzymes which increase stability may be detrimental to function (BEADLE and SHOICHET 2002; SHOICHET *et al.* 1995) are often quoted. Apparently, the first argument is circular, as pointed out by Wilke and coauthors (BLOOM *et al.* 2007). The flaw in the second (''experimental'') argument is that it fails to recognize that in many experiments which are often quoted in support of stability-function tradeoff, mutations are introduced in active sites only (BEADLE and SHOICHET 2002; SHOICHET *et al.* 1995).

In reality, upon thermal adaptation, stabilizing mutations can be introduced anywhere in the protein rather than in its active site only and those mutations, while stabilizing proteins, do not compromise, in most cases, their catalytic activity (BLOOM *et al.* 2006; GIVER *et al.* 1998). Furthermore Arnold and coauthors showed for a large number of random mutations in a mesophilic enzyme esterase that stability and catalytic activity are not inversely correlated (GIVER *et al.* 1998). The physical reason for this experimental observation is simple: even in cases when some degree of flexibility is needed for function, it requires specific local small-scale motions (like rotation of a few side-chains or loop opening and closure) which are most likely to be unrelated to motions which accompany global unfolding. Residues which are responsible for maintaining function-related flexibility are often found to be different from the ones that contribute most to protein stability (MIRNY and SHAKHNOVICH 1999). Our model reproduces distribution of protein stabilities and most phenomenology of thermal adaptation without assuming any functional penalty for protein stabilization which suggests that the stability-function trade off may not be a determining factor in thermal adaptation.

While not assuming any particular stability-function relationship our model is not neutral with respect to protein stability either, in contrast to some earlier studies (BLOOM *et al.* 2007). In addition to the assumption that unfolding of an essential protein leads to lethal phenotype (ZELDOVICH *et al.* 2007b) we also posit here that stability of rate determining proteins affects fitness through modulating the copy number of *folded* (and therefore functional) proteins as given by Eqs 1 and 2. On the other hand our model does not assume any effect of stability on catalytic rate or other functional measure of the folded protein.

It has been found that in prokaryotes denaturation temperature for certain protein families is correlated to their OGT (LI *et al.* 1998; PERL *et al.* 1998; TOPPING and GLOSS 2004). Our model provides further insight into the relation between protein stability and the OGT. Long-term adaptation upon moderate increase of temperature (up to 5C in our simulations and in many experiments) changes the distribution of protein stabilities in a strain $P(\Delta G)$ by mostly affecting least stable proteins while leaving more stable proteins whose denaturation temperature is considerably above new evolutionary temperature (and therefore are not ''threatened'' by temperature increase) relatively unchanged. Figure 3 as well as Table 1 show that adaptation affects mostly relatively unstable proteins in bacteria (by lifting the LDT significantly) while keeping MDT averaged over all proteins relatively unchanged. It is noteworthy that a few cases where correlation between stability and OGT was documented it concerned proteins which are relatively unstable in mesophilic species (LI *et al.* 1998; PERL *et al.* 1998; TOPPING and GLOSS 2004). While this coincidence is suggestive, more systematic studies are needed to confirm or falsify this prediction form our model.

From the numerical study, we have also seen how the fitness of bacteria and viruses shows different thermal responses, even for the same number of rate determining genes and the same metabolic reaction free energy barrier. This is primarily because high mutation rate in RNA viruses results in less stable proteins than those of DNA-based organisms as suggested also in recent empirical observations (TOKURIKI *et al.* 2009). Thus, when temperature is increased, a greater fraction of rate determining proteins in the RNA genome organism would unfold, and the fitness is affected more. Given the same number of rate determining genes, the fitness of RNA viruses would be affected more drastically by increasing temperature.

However, compared to DNA based organisms, RNA viruses have much shorter genomes. Bacteria usually have several hundred essential genes, while RNA viruses usually have less than 20 proteins in total. From our analysis, the more rate determining genes an organism has, the more marginally-stable-proteins this organism will have. Therefore, although RNA viruses have less stable proteins on average, their small genome size gives them fewer marginally stable proteins. This is the reason why experimentally, RNA viruses and bacteria both show significantly reduced fitness upon increasing temperature.

On the other hand, because of their high mutation rate, RNA viruses also evolve faster and are more adaptable. Therefore, after evolving in the steady environment for the same number of generations, the RNA-based organisms would likely have a higher fitness than those of the DNA-based organisms, since there have been more mutations and thus more opportunities for the proteins to re-equilibrate.

We can observe these effects on Figure 2. Initially, bacterial proteins are more stable the RNA virus ones, as shown in figure 2(B) and (C). However, since the bacterium in our model has more rate determining genes (n = 50 as in the case of Figure 2) than RNA virus (n = 20), their instant thermal response for the wild-type strain follows similar patterns (as shown in Figure 2(A)). Both bacteria and the RNA virus have an OGT relatively close to its evolutionary temperature, and both exhibit a sharp fitness decline upon temperature increase. However, after evolving at elevated environmental temperature (42°C) for 10,000 generations, the RNA virus has greatly improved its fitness in a broad temperature range including original evolutionary temperature of 37 C, as the maximum growth rate can reach up to 1.8 times of the wild-type growth rate at the original temperature, 37°C. The bacterium cultured at 42°C for same 10,000 generations has made limited improvement compared to the RNA virus; the adapted bacterium's maximum growth rate is around 1.2 times of the natural growth rate. This is in agreement with previous experimental results, which showed that some viruses can adapt to different environments within a short amount of time (HOLDER and BULL 2001; KNIES *et al.* 2006); while as after several years of evolution, the relative fitness of different *E. coli.* strains have only limited change, from 0.8 to 1.2, depending on the specific growth conditions and initial strains (BENNETT and LENSKI 2007). Experimentally, it has been found that some specific viruses can adapt to higher temperatures in a relatively short time (400 generations for bacteria phage G4, for example (KNIES *et al.* 2006)). For a bacterium with life cycle of around four hours, as in Lenski's previous experiments (they estimated around $\log_2 100 \sim 6.6$ generations of *E. coli.* per day) (COOPER *et al.* 2001), ten thousand generations is roughly five years. For *E. coli.* growing in more optimal conditions, the life cycle can be around one hour (WOLDRINGH 1976), and five thousand generations would be around 200 days. Therefore, host organism responses such as fever are effective methods to combat most bacterial infections. For viral infections, because some virus can adapt to novel thermal environment in as short as 400 generations (KNIES *et al.* 2006), given a generation time of a couple of hours, this means that they can adapt to novel environment in several days to a few months. Thus according to our model, fever might not always be the most effective mechanism for viral infections all the time, although it still may be effective to some viral infections. This agrees with common knowledge that fever response is more often caused by bacterial infection, since during the time course of a fever, which is around a few days, bacteria can hardly adjust to the new thermal environment, and are therefore likely to get eliminated.

## METHODS

As described in the main text, an organism can duplicate under certain replication and mutation rate parameters that are determined by its genotypic features such as protein stabilities and the number of rate determining genes in the organism.

In the numerical algorithm, we first prepare initial species with 1000 identical organisms of the same genotype; We here define $(\Delta G_1, \Delta G_2, \ldots \Delta G_i \ldots \Delta G_n)$ as the organism's genotype, where $\Delta G_i$ represents the free energy of protein folding (i.e. the free energy difference between the folded and unfolded states) for the i$^{th}$ rate determining protein in the organism, measured in kcal/mol. Stabilities of n rate determining proteins in each organism constituting the initial population have random values drawn from the analytical distribution of the functional form described in our previous work (ZELDOVICH *et al.* 2007b). Here for the bacteria species, we assume rate determining gene number n =

50, which is roughly 18% of the number of its essential genes (FORSYTH *et al.* 2002; GERDES *et al.* 2003); and H, the metabolic reaction free energy barrier is 20 kcal/mol. We also assume the mutation rate for bacteria species as m = 0.003 mutations per genome per replication as previously studied (DRAKE *et al.* 1998), this value is still much smaller than the mutation rate of RNA viruses. For the RNA viruses, because of their short genome and high mutation rate, we take n = 20, H = 20kcal/mol, and m = 1.5 as the corresponding parameters. In both cases of bacteria and viruses the important population genetics parameter $Nm > 1$ as it is in reality.

At each time step an organism can replicate with probability determined by the genotype-dependent replication rate as given by Eq.(4). An organism is eliminated as soon as lethal mutation occurs which confers any of its proteins folding free energy value greater than zero. Upon replication, mutations may happen in a descendant organism. Mutation in our model represents the change in stability of one or more proteins in the daughter organism compared with parent organism, i.e. genotype of the daughter organism can be presented as

$$\{\Delta \vec{G}\}_{daughter} = \{\Delta \vec{G}\}_{parent} + \{\Delta \Delta \vec{G}\}$$

where $\{\Delta\Delta\vec{G}\} = (\Delta\Delta G_{i_1}, \Delta\Delta G_{i_2}....\Delta\Delta G_{i_s})$ describes changes of stabilities upon a replication event which resulted in $s$ mutations in proteins $(i_1, i_2....i_s)$. For semi-conservative replication, mutations might occur in both the parent copy and the descendent copy. If it is a conservative replication, mutations would then occur in the descendent copy only. We generate the number of mutations $s$ at each replication effect in a daughter organism, according to a Poisson distribution, and the parameter of the Poisson distribution $m_{organism}$ is the average number of mutations per genome per replication, for this particular species. The mutation rate for each gene in each copy is then $m_{gene} = m_{organism}/N$. After selecting $s$ - the total number of proteins to be mutated at a given replication event - we decide which proteins to mutate by selecting the set $(i_1, i_2 ... i_s)$ at random. When a mutation occurs in a protein, the protein's sequence would be physically changed. In this study we do not consider protein sequences explicitly. Rather we posit that free energy of the mutant protein might have a different folding free energy, and the free energy difference $\Delta\Delta G_i$ between wild-type and mutant protein is a random value drawn from a distribution based on statistics of free energy changes collected in multiple protein engineering experiments. To this end, we determine the statistics of changes of protein stability upon mutations from the ProTherm database (KUMAR *et al.* 2006). This database contains information on more than three thousand point mutations, across all currently performed point mutation experiments. The statistics show that protein folding stability change due to point mutation roughly forms a Gaussian distribution, where the mean is 1kcal/mol and the standard deviation is 1.7kcal/mol. Therefore, when mutation occurs, we alter the protein stability by an amount drawn from this Gaussian distribution. We assume that statistics of changes of protein stability $\Delta\Delta G$ does not depend on stability $\Delta G$ itself. A similar assumption was made in (BLOOM *et al.* 2007). The mutant daughter organism will therefore have an altered fitness value (derived from (Eq. (2))), due to the altered stability of some of its proteins.

We also impose an upper limit of population size of N=10000 organisms by culling excess organisms at random. This population size ensures that for both viruses and bacteria the important population genetics parameter $Nm \gg 1$. We ran many series of independent simulations to eliminate the genetic drift effect imposed by this upper limit on the total population. During the numerical simulation, we let organisms evolve in a stable environment, for around 20000 generations, and we study the population dynamics and evolution of protein stabilities under various parameters. Parameter $b_o$ establishes the correspondence between ral time and time step in the simulation. We chose to ensure that for the initial replication rate each organism has 0.1 probability of replication at each time step.

After evolving the organisms, according to the above protocols, for 20,000 generations we first checked that distribution of protein stabilities in the populations equilibrated. To this end we checked that it did not change further after 20,000 generations of evolution at constant temperature, which was indeed the case (data not shown). Next, we performed the thermal adaptation simulations. We first study the instantaneous thermal response of replication rate to abrupt temperature changes. At this relatively short time scale, the organism's protein stability distribution would not have time to re-adjust to the environment. Each of the rate determining proteins will become less stable by $k_B \overline{\Delta S} * \delta T$, therefore decreasing the fraction of properly folded ones in the organism. For example, in order to obtain the fitness of a model organism at elevated temperature, such as $42^oC$, we can take the well evolved (at 37C) population, change the environmental temperature to $42^oC$, then recalculate the stability of each protein at this elevated environmental temperature.

In addition to instantaneous thermal response, we also study the long-time adaptation of bacteria and viruses. Here we take the evolved bacteria and virus strain (both evolved in $37^oC$ environment for 20000 generations). We increase temperature to $42^oC$, let the species evolve in the new thermal environment for 10000 generations, which for a growth cycle of ~6.6 generations per day (COOPER *et al.* 2001), is around five years, as experimentally this have been done. Afterwards, study the fitness of the organism at various temperature shifts from the $37^oC$ environment. We measure temperature also by $\delta T$, which denotes the shift from the original evolutionary temperature of $37^oC$, and we measure the *relative* fitness as a ratio of the adapted strain's growth rate to the original wild-type growth rate.

# SUPPLEMENTARY INFORMATION

## A. Derivation of Various Coefficients

According to the analysis in the main text, at instant increase of temperature to $T+\delta T$, for $-L+\overline{\Delta S}k_B\delta T < \tilde{G}_i < \overline{\Delta S}k_B\delta T$, we have the relationship of free energy change and free energy distribution change as:

$$\begin{cases} \tilde{G}_i = G_i + \overline{\Delta S}k_B\delta T \\ p(\tilde{G}_i) = C_0(-e^{\frac{\tilde{G}_i - \overline{\Delta S}k_B\delta T}{D}} Sin[\pi \frac{\tilde{G}_i - \overline{\Delta S}k_B\delta T}{L}]) \end{cases}$$

The birth rate at temperature $T+\delta T$ can therefore be expressed as:

$$<\ln b(T+\delta T)> = \ln b_0 - \frac{H}{k_B(T+\delta T)} - n \int_{-L+\overline{\Delta S}k_B\delta T}^{\overline{\Delta S}k_B\delta T} \ln(1+e^{\frac{\tilde{G}_i}{k_B(T+\delta T)}}) p(\tilde{G}_i) d\tilde{G}_i$$

The first term on the right hand side is constant with respect to temperature. The second term, $-\frac{H}{k_B(T+\delta T)}$ is the metabolic reaction barrier term, and it increases upon the increase of temperature. The third term can be evaluated by perturbation, and it's behavior upon the change of temperature can be studied in this way as well.

From the integration term above, denote $y = \tilde{G}_i - \overline{\Delta S}k_B\delta T$, then the integration part changes to:

$$\int_{-L+\overline{\Delta S}k_B\delta T}^{\overline{\Delta S}k_B\delta T} \ln(1+e^{\frac{\tilde{G}_i}{k_B(T+\delta T)}}) p(\tilde{G}_i) d\tilde{G}_i = \int_{-L}^{0} \ln(1+e^{\frac{y+\overline{\Delta S}k_B\delta T}{k_B(T+\delta T)}})(-C_0)(e^{\frac{y}{\Psi}} Sin[\frac{\pi y}{L}]) dy$$

Therefore, the birth rate at temperature $\delta T + T$ can be expressed as:

$$< \ln b(T+\delta T) > = \ln b_0 - \frac{H}{k_B(T+\delta T)} - n\int_{-L}^{0} \ln(1+e^{\frac{G+\overline{\Delta S}k_B\delta T}{k_B(T+\delta T)}})p(G)dG \quad (1)$$

Denote the third term as:

$$I(T+\delta T) = -n\int_{-L}^{0} \ln(1+e^{\frac{G+\overline{\Delta S}k_B\delta T}{k_B(T+\delta T)}})p(G)dG$$

The physical meaning of this term stands for the average logarithm concentration of the folded protein under temperature $T+\delta T$. $\ln(1+e^{\frac{G+\overline{\Delta S}k_B\delta T}{k_B(T+\delta T)}})$ can be expanded as:

$$\ln(1+e^{\frac{G+\overline{\Delta S}k_B\delta T}{k_B(T+\delta T)}}) = -\frac{\delta T}{k_B T^2}f_1(T) + \frac{1}{2}(\frac{\delta T}{k_B T^2})^2 f_2(T),$$

where:

$$\begin{cases} f_1(T) = \frac{e^{\frac{G}{k_B T}}}{(1+e^{\frac{G}{k_B T}})}(G - \overline{\Delta S}k_B T) \\ f_2(T) = \frac{e^{\frac{G}{k_B T}}}{(1+e^{\frac{G}{k_B T}})^2}(G - \overline{\Delta S}k_B T)(G - \overline{\Delta S}k_B T + 2k_B T *(1+e^{\frac{G}{k_B T}})) \end{cases} \quad (2)$$

We can express the log ratio of folded protein at $T+\delta T$ as:

$$\frac{I(T+\delta T)}{n} = \frac{\delta T}{k_B T^2}\int_{-L}^{0} f_1(T)p(G)dG - \frac{1}{2}(\frac{\delta T}{k_B T^2})^2\int_{-L}^{0} f_2(T)p(G)dG \quad (3)$$

This integral, although possible to evaluate fully analytically for both the first and the second order term, the complete integration result is lengthy. However, since for mesophiles, T~300K, we have the relationship $L >> k_B T$, $\overline{\Delta S}k_B * T >> G$. Therefore we can approximate (2) as to:

$$\begin{cases} f_1(T) = -\overline{\Delta S}k_B T e^{\frac{G}{k_B T}} \\ f_2(T) = (-\overline{\Delta S}k_B T)^2 e^{\frac{G}{k_B T}} \end{cases} \quad (4)$$

This is a great simplification. Evaluating (3) is straightforward after employing the simplified functional form of (4), noticing $e^{-\frac{L}{k_B T}} -> 0, e^{-\frac{L}{\Psi}} -> 0,$ the integration result can be further approximately expressed as:

$$\int_{-L}^{0} f_1(T) p(G) dG = \frac{\overline{\Delta S} k_B T * (k_B T)^2}{(\Psi^2 + (k_B T)^2)}, \quad \int_{-L}^{0} f_2(T) p(G) dG = -\frac{(\overline{\Delta S} k_B T)^2 * (k_B T)^2}{(\Psi^2 + (k_B T)^2)}$$

According to the analysis above, we can express the logarithmic ratio of birth rate as:

$$< \ln(\frac{b(T+\delta T)}{b(T)}) >= \delta T (n C_1 + C_2) + \delta T^2 (n C_3 + C_4),$$

where the coefficients are

$$\begin{cases} C_1 \sim -\frac{\overline{\Delta S} k_B * k_B T}{(\Psi^2 + (k_B T)^2)} \\ C_2 = \frac{H}{k_B T^2} \\ C_3 \sim -\frac{(\overline{\Delta S} k_B)^2}{2(\Psi^2 + (k_B T)^2)} \\ C_4 = -\frac{H}{k_B T^3} \end{cases}.$$

These values can then be used in analytical discussion for how the metabolic reaction free energy barrier and rate determining gene number in each species affect their thermal response behavior, as explained in the main text.

**B. Information for all 35 Datasets**

Here we provide a table for all 35 mesophilic bacteria whose thermal response has been studied (RATKOWSKY *et al.* 2005). The correlation between experimental fit and our theoretical prediction are from 90% to 99%.

|  | H | n | correlation |
|---|---|---|---|
| lmono 1 | 7.6483 | 24 | 97.88% |
| lmono2 | 6.608 | 22 | 96.08% |
| lmono 3 | 7.3372 | 24 | 96.68% |
| lmono 4 | 7.6 | 38 | 97.75% |
| lmono 5 | 7.2874 | 35 | 97.48% |
| g punicea | 7.2387 | 11 | 96.43% |
| galidibacter | 4.8427 | 11 | 89.96% |
| shewanella | 7.3522 | 37 | 96.71% |
| shewanella | 7.9265 | 22 | 99.33% |
| shewanella | 6.0716 | 20 | 97.88% |
| a. hydrophila | 8.5582 | 14 | 99.16% |
| l mono scott | 8.4017 | 18 | 98.84% |
| e coli m23 | 10.5478 | 19 | 97.77% |
| ps florescence 1412 | 8.0594 | 34 | 98.76% |
| kleb oxy | 8.1544 | 19 | 98.39% |
| p putida 1412 | 6.9578 | 36 | 97.57% |
| K120-6 | 11.8016 | 32 | 93.07% |
| K118-4 | 11.1957 | 46 | 98.48% |
| BC-29 (Exp. 3) | 9.2917 | 32 | 96.92% |
| BC-29 (Exp. 2) | 9.4541 | 42 | 98.75% |
| BC-29 (Exp. 1) | 11.102 | 41 | 98.18% |
| BC-14 (Exp. 5) | 9.4761 | 13 | 97.53% |
| BC-14 (Exp. 3) | 10.0644 | 32 | 99.01% |
| BC-14 (Exp. 2) | 7.2467 | 12 | 98.88% |
| BC-14 (Exp. 1) | 10.729 | 31 | 98.73% |
| BC-14 (Exp. 3) | 8.8185 | 18 | 97.26% |
| E.coli ONT H8 (R91) Mark Salter's thesis: | 8.6747 | 27 | 98.18% |
| E.coli O126:H21(R10) Mark Salter's thesis: | 7.6515 | 24 | 98.10% |
| E.coli NT (R31) Mark Salter's thesis: | 8.5031 | 45 | 97.24% |
| O81:H- (R106) | 8.8661 | 46 | 97.80% |
| O88:H- (R171) | 8.8661 | 46 | 97.80% |
| O88:H- (R172) | 8.7549 | 48 | 97.60% |
| O157:H- | 9.0366 | 40 | 97.75% |
| O157:H7 (EH9) | 8.8526 | 19 | 98.04% |
| O111:H- | 8.6179 | 20 | 98.50% |